\documentclass[titlepage,pallis,twoside]{wpallis}
\usepackage{fancyh}
\usepackage[l]{floatflt}
\usepackage{epsfig}
\usepackage{amssymb}
\usepackage{latexsym}
\usepackage{times}
\usepackage{slashed,cite}
\usepackage{upgreek}
\numberwithin{equation}{section}
\usepackage{rotating}
\usepackage{amsmath}
\usepackage{amsfonts}
\usepackage{amsbsy}
\usepackage{amscd}
\usepackage{bbm}
\usepackage{float}
\usepackage{bm}

\usepackage{graphics}

\newcommand{\bal}{\begin{align}}
\newcommand{\eal}{\end{align}}
\newcommand{\beqs}{\begin{subequations}}
\newcommand{\eeqs}{\end{subequations}}
\newcommand{\eec}{\end{center}}
\newcommand{\bec}{\begin{center}}

\newcommand{\ecs}{\end{cases}}
\newcommand{\bcs}{\begin{cases}}

\newcommand{\eem}{\end{matrix}}
\newcommand{\bem}{\begin{matrix}}
\newcommand{\eeq}{\end{equation}}
\newcommand{\beq}{\begin{equation}}
\newcommand{\ba}{\begin{array}}
\newcommand{\ea}{\end{array}}
\newcommand{\bea}{\begin{eqnarray}}
\newcommand{\eea}{\end{eqnarray}}
\newcommand{\baq}{\begin{eqnarray}}
\newcommand{\eaq}{\end{eqnarray}}

\newcommand\eqs[2]{Eqs.~(\ref{#1}) and (\ref{#2})}

\newcommand\eqss[3]{Eqs.~(\ref{#1}), (\ref{#2}) and (\ref{#3})}

\newcommand{\ftn}{\footnotesize}

\newcommand{\GeV}{{\mbox{\rm GeV}}}

\newcommand{\etal}{{\it et al.\/}}

\def\to{\rightarrow}

\def\lf{\left(}
\def\rg{\right)}

\newcommand\vevi[1]{\langle {#1} \rangle_{\rm I}}

\newcommand{\Vhi}{\ensuremath{ V_{\rm I}}}
\newcommand{\Vjhi}{\ensuremath{V_{\rm I}}}
\newcommand{\Hhi}{\ensuremath{ H_{\rm I}}}

\newcommand{\Khi}{\ensuremath{K}}
\newcommand{\Whi}{\ensuremath{W}}
\newcommand{\Vhio}{\ensuremath{V_{\rm I0}}}
\newcommand{\Ns}{\ensuremath{{N_\star}}}

\newcommand{\mP}{\ensuremath{m_{\rm P}}}

\def\openone{\leavevmode\hbox{\small1\kern-3.8pt\normalsize1}}
\newcommand{\dV}{\ensuremath{\Delta V_{\rm I}}}

\newcommand{\fr}{\ensuremath{j_{l{\rm M}}}}
\newcommand{\frs}{\ensuremath{j_{l{\rm M}\star}}}
\newcommand{\far}{\ensuremath{j_{1{\rm M}}}}

\newcommand{\fbrs}{\ensuremath{j_{2{\rm M}\star}}}

\newcommand{\fm}{\ensuremath{f_{\rm M}}}
\newcommand{\fb}{\ensuremath{f_{\rm T}}}

\newcommand{\fms}{\ensuremath{f_{\rm M\star}}}

\newcommand{\kst}{\ensuremath{K_{\rm st}}}

\newcommand{\kle}{\ensuremath{K_{l\rm E}}}
\newcommand{\kbe}{\ensuremath{K_{\rm 2E}}}
\newcommand{\kae}{\ensuremath{K_{\rm 1E}}}

\newcommand{\tkbse}{\ensuremath{\widetilde K_{\rm 2Es}}}
\newcommand{\tkase}{\ensuremath{\widetilde K_{\rm 1Es}}}
\newcommand{\klt}{\ensuremath{K_{l\rm T}}}
\newcommand{\kbt}{\ensuremath{K_{\rm 2T}}}
\newcommand{\kat}{\ensuremath{K_{\rm 1T}}}
\newcommand{\kam}{\ensuremath{K_{\rm 1M}}}
\newcommand{\kbm}{\ensuremath{K_{\rm 2M}}}

\newcommand{\tkbst}{\ensuremath{\widetilde K_{\rm 2Ts}}}
\newcommand{\tkast}{\ensuremath{\widetilde K_{\rm 1Ts}}}

\newcommand{\tklmst}{\ensuremath{\widetilde K_{l\rm Ms}}}
\newcommand{\tklm}{\ensuremath{\widetilde K_{l\rm M}}}
\newcommand{\klm}{\ensuremath{K_{l\rm M}}}
\newcommand{\tkamst}{\ensuremath{\widetilde K_{1\rm Ms}}}

\newcommand{\tkbmst}{\ensuremath{\widetilde K_{2\rm Ms}}}

\newcommand{\nst}{\ensuremath{N_{\rm st}}}
\newcommand{\nb}{\ensuremath{N}}
\newcommand{\na}{\ensuremath{N}}

\newcommand{\ks}{\ensuremath{k_\star}}
\newcommand{\ns}{\ensuremath{n_{\rm s}}}
\newcommand{\as}{\ensuremath{a_{\rm s}}}
\newcommand{\As}{\ensuremath{A_{\rm s}}}

\newcommand{\rcc}{\ensuremath{\mathcal{R}}}

\newcommand{\Ve}{\ensuremath{ V}}
\newcommand{\Vf}{\ensuremath{ V_{\rm F}}}

\newcommand{\what}{\ensuremath{\widehat}}

\def\bbet{{\bar\beta}}
\def\al{{\alpha}}

\def\t{\rm T}
\def\e{\rm E}
\def\m{\rm M}
\def\bpat{A$_{\rm T}$}
\def\bpae{A$_{\rm E}$}
\def\bpbt{B$_{\rm T}$}
\def\bpbe{B$_{\rm E}$}

\def\th{{\theta}}

\newcommand{\Trh}{\ensuremath{T_{\rm rh}}}
\newcommand{\sg}{\ensuremath{\phi}}

\newcommand{\sgx}{\ensuremath{\phi_\star}}
\newcommand{\sgf}{\ensuremath{\phi_{\rm f}}}

\newcommand{\ld}{\ensuremath{\lambda}}

\newcommand{\se}{\ensuremath{\widehat \phi}}
\newcommand{\sex}{\ensuremath{\widehat{\phi}_\star}}
\newcommand{\sef}{\ensuremath{\widehat{\phi}_{\rm f}}}

\newcommand{\eph}{\ensuremath{\epsilon}}
\newcommand{\ith}{\ensuremath{\eta}}

\newcommand{\Dex}{\ensuremath{\Delta_{\star}}}

\newcommand{\phc}{\ensuremath{\Phi}}
\newcommand{\phcb}{\ensuremath{\Phi^*}}

\newcommand{\nm}{\ensuremath{q_{\rm M}}}

\def\Ka{K\"{a}hler potential}
\def\Km{K\"{a}hler manifold}
\def\Kaa{K\"{a}hler~}

\newcommand{\plk}{{\it Planck}}

\def\actc{{\sf\ftn P-ACT-LB-BK18}}

\newcommand{\tmi}{{TMI}}
\newcommand{\emi}{{EMI}}
\newcommand{\etmi}{{E/TMI}}
\def\tpmi{{T$_p$MI}}
\def\epmi{{E$_p$MI}}
\newcommand{\etpmi}{{E$_p$/T$_p$MI}}

\renewcommand{\figurename}{\bf\scshape Fig.}
\renewcommand{\tablename}{\bf\scshape Table}
\renewcommand{\refname}{{\bf\scshape References}}

\begin{document}

%\preprint{UT-STPD-2/10} Pole Inflationin view of ACT $(N,p)$

\title{\boldmath\bf\scshape ACT-Inspired K\"ahler-Based Inflationary Attractors}

\author{\large\bfseries\scshape  C. Pallis}
\address[] {\sl  School of Technology,  \\ Aristotle
University of Thessaloniki, \\ Thessaloniki, GR-541 24 GREECE \\
{\sl e-mail address: }{\ftn\tt kpallis@auth.gr}}

%\date{\today}

%of the recent proposed models of inflation

\begin{abstract}

\noindent {\small \bf\scshape Abstract} \\ \par We develop a new
class of cosmological attractors which are compatible with the
recent ACT results. They are based on two types of fractional \Ka
s, $K$, for a gauge-singlet inflaton $\sg$ which reduce, along the
inflationary path, to the form $N/(1-\sg^{\nm})^{p}$ with $\nm=1,
2$ and $0.1\leq p\leq10$. The combination of these $K$'s with the
chaotic potentials $\phi^n$ (where $n=2, 4$) within a non-linear
sigma model leads to inflationary observables which are consistent
with the current data and largely independent from $\nm$ and $n$.
Endowing these $K$'s with a shift symmetry we also offer a
supergravity realization of our models introducing two chiral
superfields and a monomial superpotential, linear with respect to
the inflaton-accompanying field. The attainment of inflation with
subplanckian inflaton values and the large values for the
tensor-to-scalar ratio, which increases with $N$, are two
additional attractive features of our proposal.
\\ \\ {\ftn {\sf PACS codes: 98.80.Cq, 11.30.Qc, 12.60.Jv,
04.65.+e}\\ {\sf Keywords: Cosmology, Supersymmetric models,
Supergravity}}\\ [0.2cm] \publishedin{{\sl  J. Cosmol. Astropart.
Phys.} {\bf 09}, 061 (2025)}
\end{abstract} \maketitle
%https://doi.org/10.1088/1475-7516/2025/09/061

\pagestyle{fancyplain}

\maketitle

\rhead[\fancyplain{}{ \bf \thepage}]{\fancyplain{}{\sl
ACT-Inspired K\"ahler-Based Inflationary Attractors}}
\lhead[\fancyplain{}{\sl C. Pallis}]{\fancyplain{}{\bf \thepage}}
\cfoot{}

\tableofcontents\vskip-1.3cm\noindent\rule\textwidth{.4pt}\\
%\vspace*{.3cm}

\section{Introduction} \label{intro}

The presence of a pole
\cite{terada,pole,pole1,sor,polec,epole,ethi,tmhi,lee} in the
kinetic term of the inflaton, $\sg$, with chaotic potential
$\sg^n$ gives rise to inflationary models collectively named
$\alpha$ (or $N$ in our notation) attractors \cite{alinde, prl}.
Prominent representatives are the \emph{E-Model Inflation}
({\sf\ftn EMI}) \cite{alinde} (or $\alpha$-Starobinsky model
\cite{eno7}) and \emph{T-Model Inflation} ({\sf\ftn \tmi})
\cite{tmodel} which assure (independently of $n$) a
tensor-to-scalar ratio $r$ increasing with $N$ and a (scalar)
spectral index \ns\ consistent with \plk\ data \cite{plin}, i.e.,
\beq \label{nsnmi}
\ns\simeq1-2/\Ns=0.965\>\>\mbox{where}\>\>\Ns=55 \eeq
is the number e-foldings elapsed from the crossing of the pivot
scale of the inflationary horizon until the end of inflation.
However, the value in \Eref{nsnmi} is substantially lower than the
latest \emph{Data Release 6} ({\sf\ftn DR6}) from the
\emph{Atacama Cosmology Telescope} ({\sf\ftn ACT})
\cite{act,actin}, combined with the \emph{cosmic microwave
background} ({\sf\ftn CMB} measurements by \plk\ \cite{plin} and
\emph{BICEP/Keck} ({\sf\ftn BK}) \cite{bcp}, together with the
\emph{Dark Energy Spectroscopic Instrument} ({\sf\ftn DESI}) {\it
Baryon Acoustic Oscillation} ({\sf\ftn BAO}) results \cite{desi}.
Indeed, the so-called \actc\ data entails \cite{actin}
\beq \label{data}
\ns=0.9743\pm0.0068\>\Rightarrow\>0.967\lesssim\ns\lesssim0.981\>\>\mbox{and}\>\>r\leq0.038
\eeq
at 95$\%$ \emph{confidence level} ({\sf\small c.l.}) -- the
running of $\ns$, $\as$, may be negligibly low at 95$\%$ c.l. too.
Several modifications have been recently proposed in order to
reconcile \emph{E- and T-model inflation} ({\sf\ftn \etmi}) with
data \cite{indi,act2,act4,act5,gup,oxf,kina,actlee,etwarm,ketovas}
-- similar attempts have been also performed for (non-)minimal
\cite{actattr,actlinde,aoki,maity,rhb,nmact,yin,warm,rhc,rha,rhd,act1,act3,act6,actj,actpal},
Starobinsky
\cite{actellis,r2drees,ketov,acttamv,r2a,r2b,r2mans,r2li} or
F-term hybrid \cite{fhi1,fhi2,fhi3,fhi4} inflation too.

A systematic consideration of \etmi\ -- see e.g. \cref{polec,
ethi} -- reveals that they can be identified from the order of
pole in the inflaton kinetic term. Namely, if we parameterize this
term as
\beq
\frac{N\dot\sg^2}{2\fm^2}\>\>\mbox{where}\>\>\fm=1-\sg^{\nm}\>\>\mbox{and}\>\>\nm=\bcs1&\mbox{for
$\m=\e$,}\\ 2&\mbox{for $\m=\t$,}\ecs\label{fp}\eeq
%
%Hereafter the symbolic index \m\ takes ``values'' \e\ and \t\ for
%\emi\ and \tmi\ respectively.
then \emi\ [\tmi] occurs for \m=\e\ [\m=\t] -- here dot denotes
derivation \emph{with respect to} ({\ftn\sf w.r.t}) the cosmic
time. Both options can be produced from specific logarithmic \Ka s
\cite{epole,tmodel}.  It is noticed, however, that a kinetic term
of the form \cite{prl, terada}
\beq \frac{N\dot\sg^2}{2\fb^{\sf p}}\>\>\mbox{yields}\>\>
\ns\simeq1-\frac{\sf p}{({\sf p}-1)\Ns}.\label{fbp}\eeq
In view of \Eref{data}, we can easily deduce that the last result
modifies significantly \Eref{nsnmi} towards the correct direction
for ${\sf p}>2$ maintaining the independence from $n$ and
approximately \cite{terada} from $N$. It would be useful,
therefore, to investigate if the kinetic mixing in \Eref{fbp} can
be derived from some \Ka s. To our knowledge, this issue has not
been addressed so far, despite the fact that modifications to the
results of E/TMI due to rational corrections to their \Ka s have
been already investigated \cite{pole1,fibre,louis,all}. Although
without direct motivation from string theory, we here introduce
novel, totally rational \Ka s $\klm$ with $l=1,2$ and $\m=\e$ and
\t, which reduce along the inflationary trough to the form
$N/\fm^p$ with $\fm$ being defined in \Eref{fp}. The cooperation
of $\klm$ with the well-motivated chaotic potentials $\sg^n$ with
$n=2$ and $4$ result to {\it identical} $\ns$ values -- at lowest
order in the expansion $1/\Ns$ -- with those in \Eref{fbp}
independently from the $\nm$ and $n$ values. Therefore, our models
exhibit an attractor mechanism, as \etmi, depended exclusively on
one extra parameter $p$. For this reason we call them $(N,p)$
attractors or \emph{E$_p$- and T$_p$-Model inflation} ({\ftn\sf
\etpmi}). Apart from the increase of $\ns$ w.r.t its value in
\etmi, \etpmi\ assures also a progressive augmentation of $r$ with
$N$ which can be hopefully tested in the near future \cite{det}.

The relevant particle models can be established, in the non-SUSY
regime, as non-linear sigma models -- see \Sref{set}. Adapting
appropriately \cite{rube,su11,unvr2} these \Ka s, the same models
can be incarnated in the context of \emph{Supergravity} ({\sf\ftn
SUGRA}) -- see \Sref{sugra}. We then approach analytically the
inflationary dynamics in \Sref{ana} and derive our numerical
predictions in \Sref{num}. We conclude in \Sref{con}. Throughout,
we use the reduced-Planck units with $\mP=M_{\rm
P}/\sqrt{8\pi}=1$.

%they offer interesting inflationary phenomenology, as we see
%below, if

%and spontaneously offers excellent agreement differ appreciably

\section{non-SUSY Framework} \label{set}

Working in the context of a non-linear sigma model we assume that
the kinetic mixing in the moduli space of one (complex) scalar
field $\Phi$ is controlled by a metric $K_{\phc\phcb}$. The
relevant lagrangian terms are written as
\beq\label{action1} {\cal  L} = \sqrt{-\mathfrak{g}}
\left(-\frac{1}{2}\rcc +K_{\phc\phcb}
\partial_\mu \phc\partial^\mu \phcb-
V(\phc)\right), \eeq
where $\mathfrak{g}$ is the determinant of the background
Friedmann-Robertson-Walker metric $g^{\mu\nu}$ with signature
$(+,-,-,-)$, $\rcc$ is the Ricci scalar  and star ($^*$) denotes
complex conjugation. We further assume that $K_{\al\bbet}$
originates from a K\"ahler potential $K$ according to the generic
definition
\beq \label{kdef} K_{\al\bbet}={K_{,z^\al
z^{*\bbet}}}>0\>\>\>\mbox{with}\>\>\>K^{\bbet\al}K_{\al\bar
\gamma}=\delta^\bbet_{\bar\gamma},\eeq
where $z^\al$ are complex scalar fields -- the symbol $,z^\al$ as
subscript denotes derivation w.r.t the field $z^\al$. Trying to
generate kinetic mixings similar to that in \Eref{fbp} we consider
%-- which can be collectively denoted as $\klm$ with $l=1,2$ and
%\m=\e\ and \t --
four $K$'s, $\klm$ with $l=1,2$ and $\m=\e, \t$, defined for
$\na>0$ as follows
\beq\kam={\na}{\left(1-(2\phc\phcb)^{\nm/2}\right)^{-p}}\>\>\>\mbox{and}\>\>\>
\kbm={\nb}{\left(1-2^{(\nm-2)/2}(\phc^{\nm}+\phc^{*\nm})\right)^{-p}},\label{kabm}\eeq
where $\nm$ is defined in \Eref{fp}. More explicitly, $\klm$ above
are specified as follows
\beqs\beq\kae={\na}{\left(1-(2\phc\phcb)^{1/2}\right)^{-p}}\>\>\>\mbox{and}\>\>\>
\kbe={\nb}{\left(1-(\phc+\phc^{*})/\sqrt{2}\right)^{-p}}\>\>\>\mbox{for
\epmi}\label{kabe}\eeq
with $(\phc\phcb)^{1/2}:=|\phc|<1/\sqrt{2}$ and
$\phc+\phc^{*}<\sqrt{2}$ respectively and
\beq\kat={\na}{\left(1-2|\phc|^2\right)^{-p}}\>\>\>\mbox{and}\>\>\>
\kbt={\nb}{\left(1-\phc^2-\phc^{*2}\right)^{-p}}\>\>\>\mbox{for
\tpmi}\label{kab} \eeq\eeqs
with $|\phc|<1/\sqrt{2}$ and $\phc^2+\phc^{*2}<1$ respectively. We
conservatively confine ourselves in the range $0.1\leq p\leq10$.
The corresponding \Kaa metrics are found to be
%For $p=1$ the $K$'s in \Eref{kab} include just quadratic terms. We
%keep, however, the dependence on $p$ confining it conservatively
%in the range $1\leq p\leq10$.
%
\begin{align} \label{mab} K_{\phc\phcb}=pN\cdot\begin{cases}
({1+\sqrt{2}p|\phc|^2})/2\sqrt{2}{(1-\sqrt{2}|\phc|)^{p+2}}&\mbox{for $K=\kae$,}\\
(p+1)/2{(1-(\phc+\phc^*)/\sqrt{2})^{p+2}}&\mbox{for
$K=\kbe$,}\\
{2(1+2p|\phc|^2)}/{(1-2|\phc|^2)^{p+2}}&\mbox{for $K=\kat$,}\\
{4(p+1)|\phc|^2}/{(1-\phc^2-\phc^{*2})^{p+2}}&\mbox{for $K=\kbt$.}
\end{cases}\end{align}
All $\klm$ in \eqs{kabe}{kab} parameterize hyperbolic \Km s but
without constant curvatures as in the cases of \etmi\ -- cf.
\cref{ethi}. Note that \kam\ enjoy a $U(1)$ symmetry whereas
$\kbm$ remain invariant under the replacement
\beq \Phi^{\nm} \to\ \Phi^{\nm}+a\>\>\>\mbox{with}\>\>\>a\in
\mathbb{R}.\label{shift2}\eeq
For $\nm=2$ this is equivalent \cite{run} with a symmetry under
hyperbolic rotations -- i.e. $SO(1,1)$. These symmetries, however,
do not determine uniquely the corresponding \Km s.
%\>\>\>\mbox{and}\>\>\>\nm=\bcs1&\mbox{for \m=\e,}\\2&\mbox{for
%\m=\t.}\ecs

%Note that the kinetic mixing in \Eref{VJe} deviates from that in
%\Eref{fbp} and so the derivation of the same $\ns$ result in
%\Eref{fbp} is highly non trivial.

%%%%%%%%%%%%%%%%%%%%%%%%%%%%%%%%%%%%%%%%%%%%%%%%%%%%%%%%%%%%%%%%%%%%
\begin{figure}[!t]%\vspace*{-.25in}
\centering\centering\hspace*{-0.3cm}
\includegraphics[width=60mm,angle=-90]{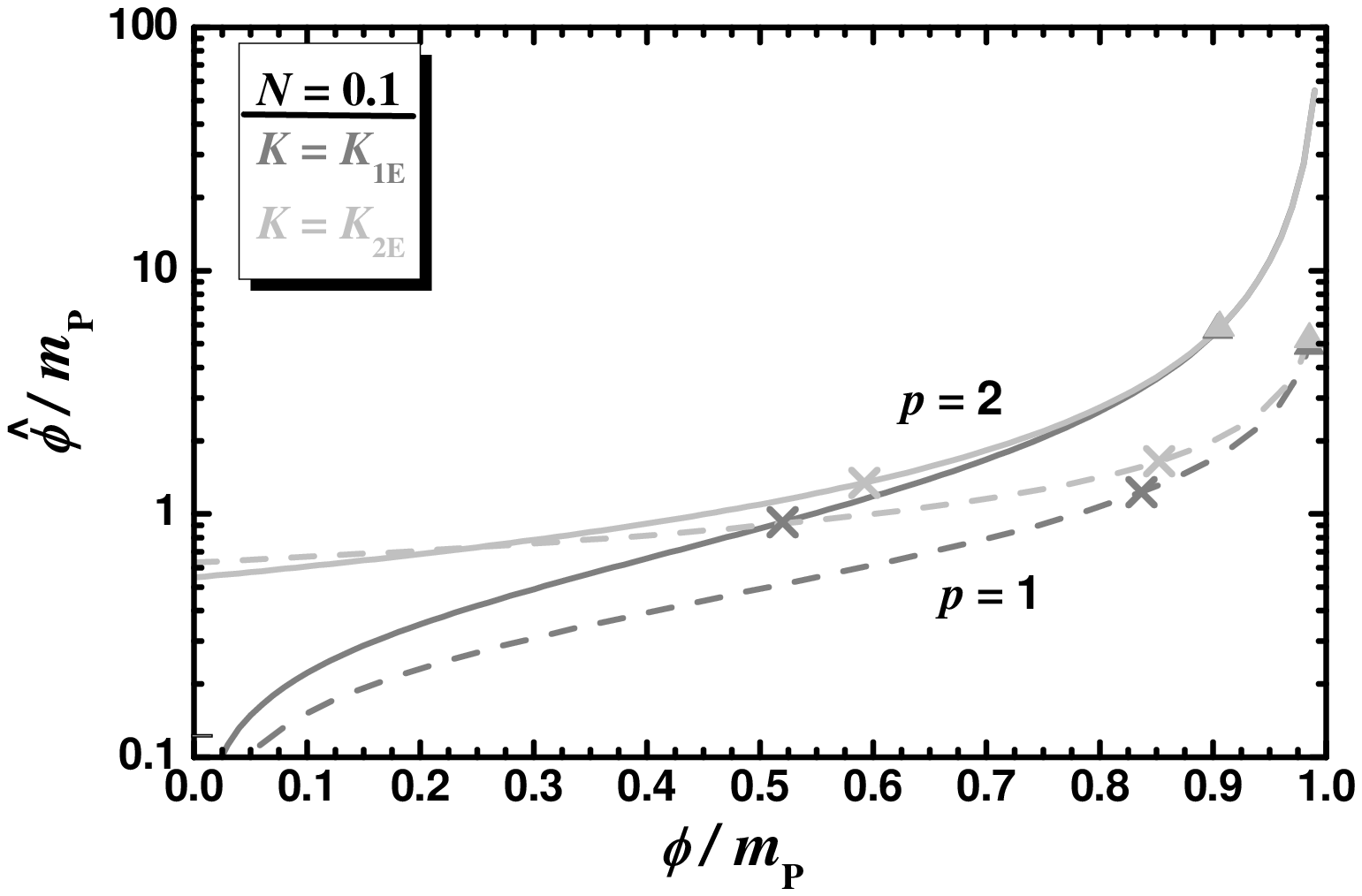}\hspace*{-0.7cm}
\includegraphics[width=60mm,angle=-90]{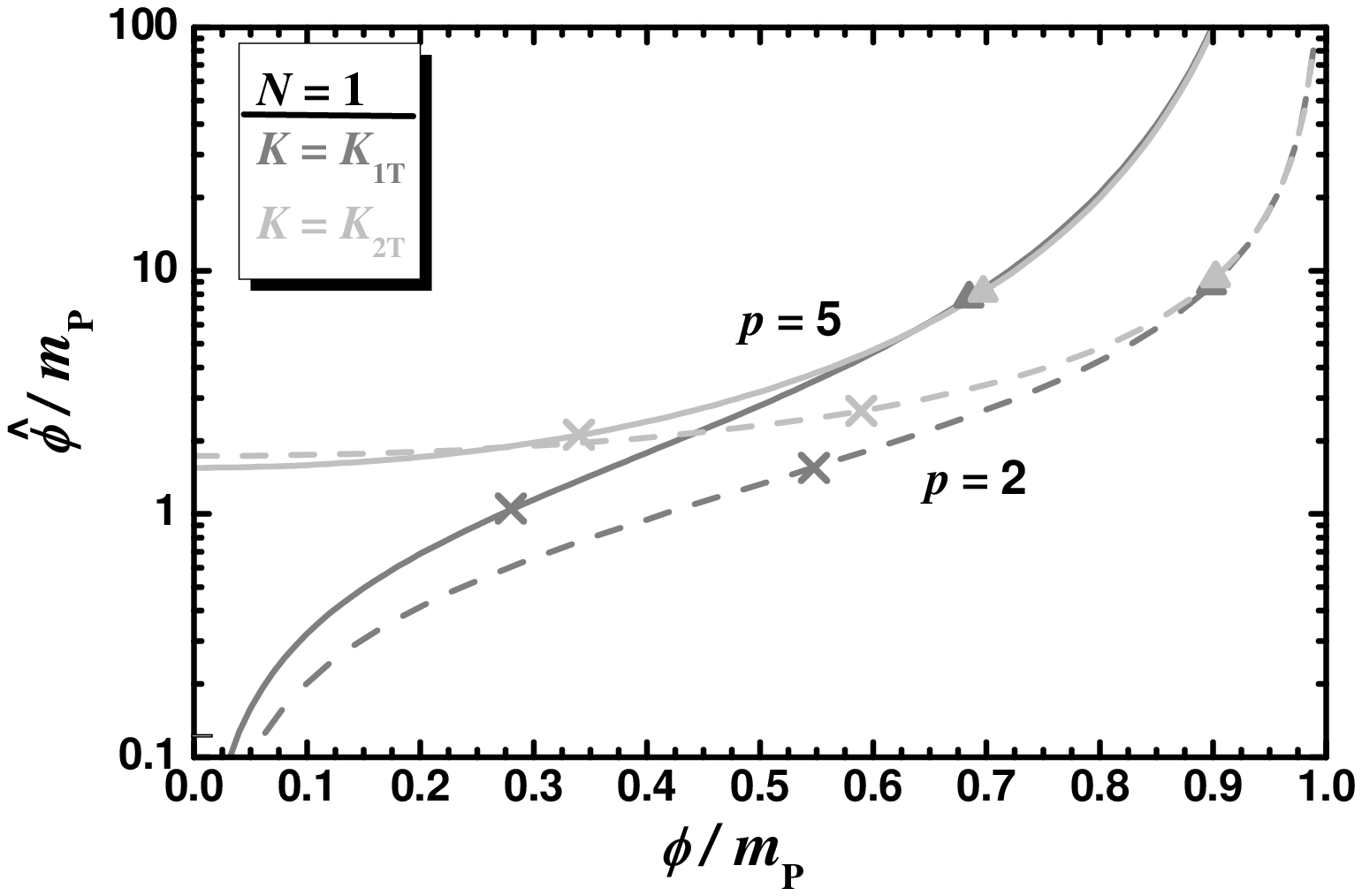}
\hfill \caption{\sl\small Canonically normalized inflaton $\se$ as
a function of $\sg$ for \epmi\ (left panel) or \tpmi\ (right
panel). We employ $K=\kae$ and \kat\ (gray lines) or $K=\kbe$ and
\kbt\ (light gray lines) for BPs of \Tref{tab1} \bpae\ and \bpat\
(solid lines) or \bpbe\ and \bpbt\ (dashed lines). }\label{fig1}
\end{figure}
%%%%%%%%%%%%%%%%%%%%%%%%%%%%%%%%%%%%%%%%%

In \Eref{action1} the potential for our models is also
incorporated. Its form is
\beq\label{vsg}
V(\phc)=\ld^2|\phc|^{n}+m^2|\phc-\phc^*|^2\>\>\mbox{with}\>\>\phc=\sg
e^{i\theta}/\sqrt{2}.\eeq
The last unusual term in $V$ is adopted to provide the angular
mode of $\phc$, $\th$, with mass, as we see below -- cf.
\cref{ethi}. $V$ in \Eref{vsg} could give rise to chaotic
inflation since it possesses a direction along which the chaotic
inflationary potentials arise. Namely, this direction is
\beq \vevi{\th}=0\>\>\mbox{with}\>\>
\Vhi:=\vevi{V}=\ld^2\sg^n/2^{n/2}, \label{vhi}\eeq
where the symbol $\vevi{Q}$ stands for the value of a quantity $Q$
during inflation. The compatibility with data is obtained thanks
to the fact that the canonically normalized inflaton $\se$ differs
from $\sg$. Indeed, it is given from the relation
\beqs\beq \label{VJe}
{d\se}/{d\sg}=J=\vevi{K_{\phc\phcb}}^{1/2}=\lf \nm^2
pN\fr^{\nm}\sg^{\nm-2}/2\fm^{p+2}\rg^{1/2}\>\>\mbox{for
$K=\klm$,}\eeq
where no summation over the repeated indices \m\ and $l$ is
applied, we retain the definition of $\fm$ in \Eref{fp} and
introduce the auxiliary function
\beq \fr=\bcs 1+p\sg^{\nm} &\mbox{for $K=\kam$ i.e. $l=1$,} \\
(1+p)\sg^{\nm} &\mbox{for $K=\kbm$ i.e. $l=2$.} \ecs \label{fr}
\eeq\eeqs
The last function assists us to present ``unified'' formulas for
all $\klm$ in \eqs{kabe}{kab}. Although the kinetic mixing in
\Eref{VJe} deviates from that in \Eref{fbp}, the resulting $\ns$
remains the same after a suitable identification of ${\sf p}$, as
we see in \Sref{ana} below.

Integrating \Eref{VJe} we can specify the functions $\se=\se(\sg)$
which have the forms
\beq\se=\sqrt{2 N p}\cdot\bcs \sg^{\nm/2} F_1\lf\frac12; 1 +
{p}/{2}, -1/2; 3/2; \sg^{\nm}, -p \sg^{\nm}\rg
&\mbox{for $K=\kam$,}\\
\sqrt{(1 + p)/p^2\fm^p}&\mbox{for $K=\kbm$.}\ecs \label{sesg} \eeq
Here $F_1(a;b_1,b_2;c;x,y)$ is the  Appell hypergeometric function
of two variables \cite{wolfram}. The numerical scrutiny of the
functions above reveals that our models exhibit a mechanism of
enhancing $\se$ w.r.t $\sg$ analogous to that obtained in the
context of E/TMI. To highlight this characteristic we first define
in \Tref{tab1} four \emph{benchmark points} ({\sf\ftn BPs})
A$_{\rm E}$, B$_{\rm E}$, A$_{\rm T}$ and B$_{\rm T}$ determined
by the variables $(n,N,p)$. We then plot in \Fref{fig1} $\se$ as a
function of $\sg$ for BPs \bpae\ and \bpbe\ with $K=\kle$ (left
plot) and \bpat\ and \bpbt\ with $K=\klt$ (right plot) -- recall
that $l=1,2$. We draw gray and light gray lines for $K=\kam$ and
$\kbm$ respectively. From the plots we can remark that $\se$
increases w.r.t $\sg$ rendering chaotic inflation possible even
for $\sg<1$. Also, for large enough $\sg$ values both $\klm$ with
fixed \m\ give almost identical $\se$ values. We depict also the
observationally relevant inflationary period which is limited
between the two $\sg$ values $\sgf$ and $\sgx$ -- see \Sref{ana}
below -- which are given in \Tref{tab1} for each BP and \klm. The
discrimination of $\sgf$ at low $\sg$ values for two different
\klm\ with fixed \m\ is negligible for the inflationary dynamics
since the observables are determined $\sg\simeq\sgx\gg\sgf$ -- see
below -- and the possible change on $\Ns$ is generically
suppressed.

%%%%%%%%%%%%%%%%%%%%%%%%%%%%%%%%%%%%%%%%%%%%%%%%%%%%%%%%%%%%%%%%%%%%
\begin{figure}[!t]\centering\hspace*{-0.3cm}
\includegraphics[width=60mm,angle=-90]{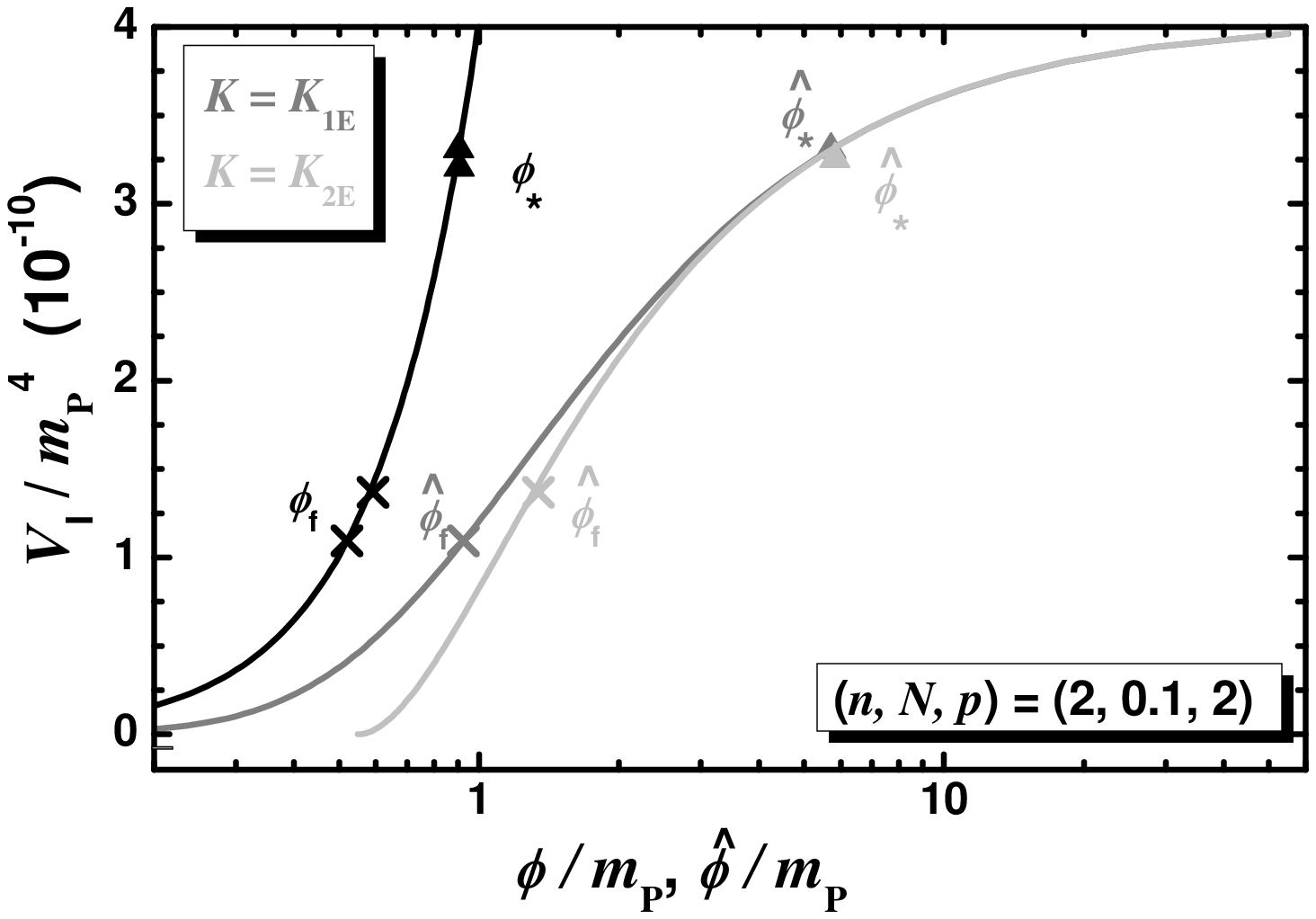}\hspace*{-0.7cm}
\includegraphics[width=60mm,angle=-90]{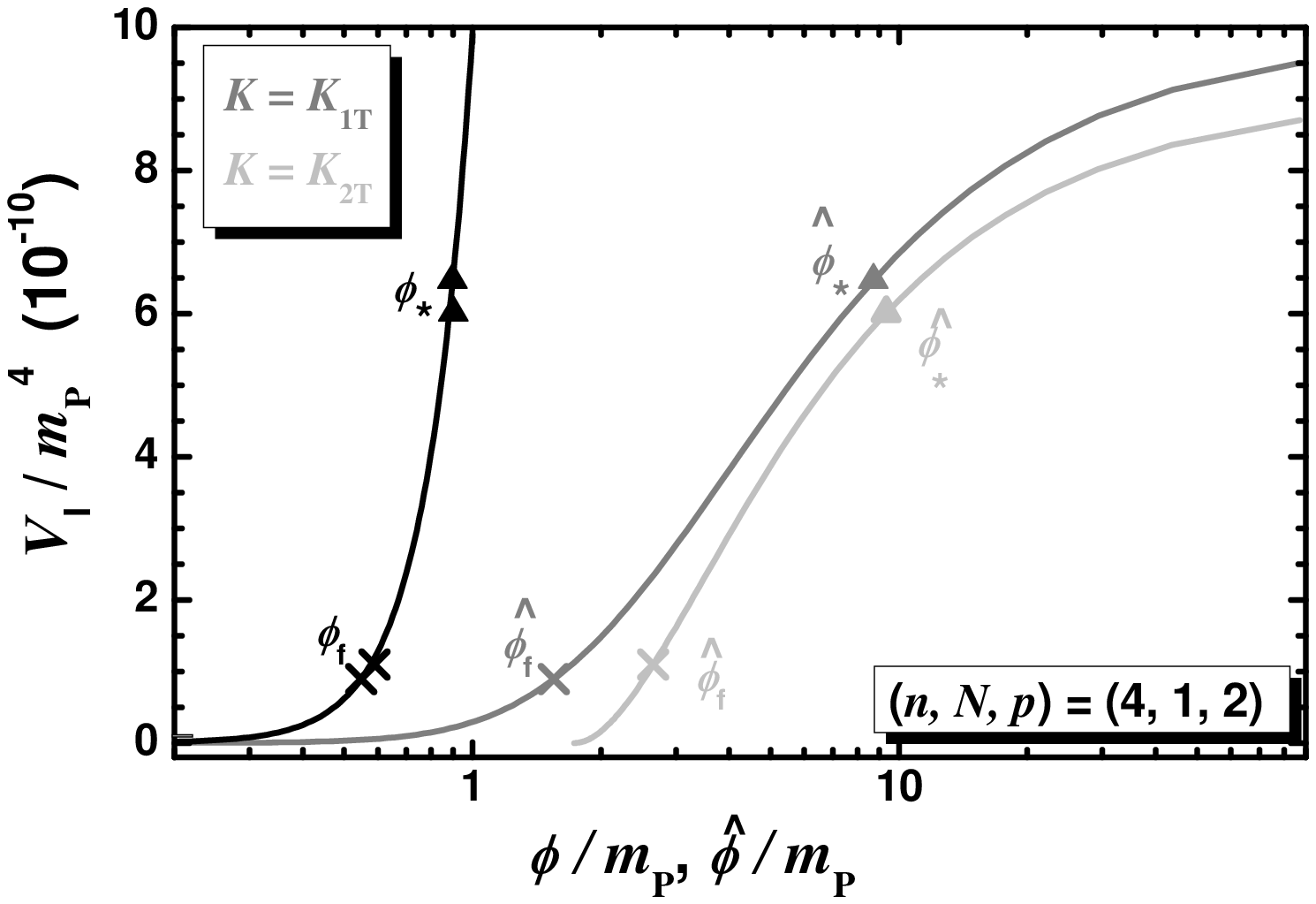}
\caption[]{\sl\small Inflationary potential $\Vhi$ for \epmi\ and
BP \bpae\ (left panel) or \tpmi\ and BP \bpbt\ (right panel) --
the BPs are defined in \Tref{tab1} -- as a function of $\sg$
(black lines) for $\sg>0$ and $\se$ for $\se>0$ and $K=\kam$ (gray
lines) or $K=\kbm$ (light gray lines) with $\m=\e$ and $\t$.
Values corresponding to $\sgx$, $\sgf$, $\se$ and $\sef$ are also
depicted. }\label{fig2}
\end{figure}
%%%%%%%%%%%%%%%%%%%%%%%%%%%%%%%%%%%%%%%%%

%
%
%\beq (n,N,p)=\bcs (2,0.1,2)&\mbox{: BP A$_{\rm E}$}\\
%(4,0.1,1)&\mbox{: BP B$_{\rm E}$}\ecs~~~~\mbox{for
%\epmi}~~\mbox{and}~~(n,N,p)=\bcs (2,1,5)&\mbox{: BP A$_{\rm T}$}\\
%(4,1,2)&\mbox{: BP B$_{\rm T}$}\ecs \mbox{for
%\tpmi}\label{bps}\eeq
%

In \etmi, the enhancing mechanism of $\se$ is accompanied with a
flattening of $\Vhi$ as a function of \se. To clarify if this is
true in \etpmi, we comparatively present in \Fref{fig2} $\Vhi$ in
\Eref{vhi} as a function of $\sg$ (black lines) and $\se$ for
$K=\kam$ (gray line) or $K=\kbm$ (light gray line). We employ BP
\bpae\ in the left plot and \bpbt\ in the right plot. The
calculation of $\Vhi$ requires the determination of $\ld$ which is
given in \Tref{tab1} and stems from the normalization of the power
spectrum of the curvature perturbation -- see \Sref{num}. We
remark that $\ld$ slightly differs for the two $\klm$'s adopted in
each plot but its impact is almost indistinguishable. In both
plots we see that $\Vhi$ as a function of $\sg$ has a
parabolic-like slope. On the contrary, $\Vhi$ as a function of
$\se$ -- for both $\klm$'s in each plot -- experiences a
stretching for $\se>1$ which results to the well-known plateau of
\etmi\ for $\se\gg1$ -- see e.g. \cref{tmodel,epole,sor,unvr2}.
Note also that $\Vhi=\Vhi(\se)$ for $K=\kbe$ and $\kbt$ is shifted
away from the origin since $\se(\sg=0)$ does not vanish. As in
\Fref{fig1}, we design the observationally relevant inflationary
period between $\sgf$ and $\sgx$ -- listed in \Tref{tab1} -- and
we show that the corresponding $\Vhi$ values for each pair of
\klm\ with fixed \m\ are very close the one to other and so we
expect identical observables.

%Let us finally remark that $\Vhi^{1/4}\ll\mP$ and so the
%semiclassical approximation, used in our analysis below, is
%perfectly valid.

% They correspond to the BPs \bpae\ for $K=\kae$ and
% $\kbe$ and \bpbt\ for $K=\kat$ and $\kbt$.

\setcounter{table}{0}
\renewcommand{\arraystretch}{1.4}
\begin{sidewaystable}[h!]\vspace*{-15.0cm}
\bec
\begin{tabular}{|c|c||c|c|c|c|c|c|c|c||c|c|c|c|c|c|c|c|}\hline
BP:&$(n,N,p)$&\multicolumn{8}{c||}{\sc Numerical
Values}&\multicolumn{8}{c|}{\sc Analytical Values}\\\cline{3-18}
&&$\ld/$&$\sgx/$&$\Dex$&$\sgf/$&$\Ns$&$\ns/$&$r/$&$-\as/$
&$\ld/$&$\sgx/$&$\Dex$&$\sgf/$&$\Ns$&$\ns/$&$r/$&$-\as/$\\
&&$10^{-5}$&$0.1$&$(\%)$&$0.1$&&$0.1$&$0.01$&$10^{-4}$

&$10^{-5}$&$0.1$&$(\%)$&$0.1$&&$0.1$&$0.01$&$10^{-4}$\\
\hline\hline
&&\multicolumn{16}{c|}{$K=\kae$ (non-SUSY Context) or $K=\tkase$
(SUGRA Context)}\\\cline{3-18}
A$_{\rm
E}$:&$(2,0.1,2)$&$2.8$&$9.04$&$9.5$&$5.2$&$51.8$&$9.73$&$1.2$&$5.3$&
$3.1$&$9.01$&$9.9$&$5.1$&$57.3$&$9.74$&$1.2$&$4.9$\\
B$_{\rm
E}$:&$(4,0.1,1)$&$2.4$&$9.85$&$1.5$&$8.4$&$55.8$&$9.73$&$0.4$&$4.9$&
$2.3$&$9.85$&$1.5$&$4.8$&$57.1$&$9.73$&$0.4$&$4.8$\\\hline
&&\multicolumn{16}{c|}{$K=\kbe$ (non-SUSY Context) or $K=\tkbse$
(SUGRA Context)}\\\cline{3-18}
A$_{\rm
E}$:&$(2,0.1,2)$&$2.8$&$9.1$&$9.4$&$5.9$&$51.4$&$9.73$&$1$&$5.3$&
$3.1$&$9.$&$9.9$&$4.9$&$60.2$&$9.74$&$1$&$5$\\
B$_{\rm
E}$:&$(4,0.1,1)$&$2.4$&$9.8$&$1.5$&$8.5$&$55.7$&$9.73$&$0.4$&$4.9$&
$2.4$&$9.85$&$1.5$&$5.6$&$57.8$&$9.73$&$0.4$&$4.8$\\ \hline
&&\multicolumn{16}{c|}{$K=\kat$ (non-SUSY Context) or $K=\tkast$
(SUGRA Context)}\\\cline{3-18}
A$_{\rm
T}$:&$(2,1,5)$&$5.7$&$6.845$&$31$&$2.8$&$51.6$&$9.73$&$2.4$&$5.4$&
$11.1$&$6.3$&$37$&$3.3$&$111$&$9.77$&$1.5$&$4.4$\\
B$_{\rm
T}$:&$(4,1,2)$&$6.3$&$8.9$&$10$&$5.5$&$56.8$&$9.74$&$2$&$4.75$&
$7.4$&$8.9$&$11$&$5.2$&$70.9$&$9.76$&$1.9$&$4.1$\\\hline
&&\multicolumn{16}{c|}{$K=\kbt$ (non-SUSY Context) or $K=\tkbst$
(SUGRA Context)}\\\cline{3-18}
A$_{\rm T}$:&$(2,1,5)$&$5.3$&$6.96$&$30$&$3.4$&$51.6$&$9.74$&$2.2$&$5.1$&
$12.5$&$6.3$&$37$&$3.2$&$135$&$9.77$&$1.5$&$4.4$\\
B$_{\rm T}$:&$(4,1,2)$&$6.$&$9.025$&$9$&$5.9$&$56.5$&$9.74$&$1.9$&$4.6$&
$7.9$&$8.9$&$11$&$4.8$&$78.6$&$9.76$&$1.9$&$4.2$\\
\hline
\end{tabular}\\[0.5cm]%\captionsetup
%{\slshape\bfseries \small Table~1:}
\captionof{table}{\sl\small
Numerical versus analytical results and comparison between the
findings for the BPs A$_{\rm M}$ and B$_{\rm M}$ using $K=\kam$ or
$\tkamst$ and $K=\kbm$ or $\tkbmst$. }\label{tab1}
\end{center}%\hspace*{5cm}\captionof{table}{} with $\m=\e$ or $\t$
\end{sidewaystable}

\clearpage
%\>\>\>\mbox{with}\>\>\>l=1,2\>\>\mbox{and}\>\>\>\m=\e,\t

We can, finally, verify that $\what{\th}= J\th\sg$ remains
well-stabilized and acquires heavy mass during \etpmi, thanks to
the last term of \Eref{vsg}. Indeed, we confirm that during \etpmi
\beq m_\th
=\frac{3\cdot2^{n/2}m^2\fm^{p+2}}{\ld^2pN\fr\sg^n}\Hhi^2\gg\Hhi^2=\frac{\Vhi}{3}\>\>\mbox{for
$K=\klm$}\label{Hhi}\eeq
provided that $m\gtrsim10^{-2.5}$. Therefore, $\what{\th}$ does
not contribute to the curvature perturbation -- see below in
\Sref{ana}. We also checked that the one-loop radiative
corrections, $\dV$, to $\Vhi$ induced by $m_{\th}$ let intact our
inflationary outputs, if we take for the renormalization-group
mass scale $Q=m_{\th}(\sgx)$ -- cf.~\cref{ethi}. E.g., for BPs
\bpae, \bpbe, \bpat\ and \bpbt\ in \Tref{tab1} and $m=10^{-2.5}$
we obtain
\beq Q=7.3\cdot10^{-5}, 2.5\cdot10^{-5},
8.5\cdot10^{-5}\>\>\mbox{and}\>\>
5.1\cdot10^{-5}\label{qnsusy}\eeq
correspondingly and using for definiteness $K=\kam$.

\section{SUGRA Framework}\label{sugra}

{Despite the fact that SUSY is not yet discovered at LHC, it
remains a viable possibility especially at high scales
\cite{susy}. It assists in stabilizing the separation of scales,
achieving precise gauge-coupling unification and providing a
cold-dark-matter candidate \cite{wells}. Therefore, it is
certainly important to investigate whether our set-up admits a
SUSY (more precisely SUGRA) incarnation.}

To achieve this goal we apply the recipes of \cref{rube, su11,
unvr2}. In particular, we consider two gauge-singlet chiral
superfields, i.e., $z^\al=\Phi, S$, with $\Phi$ ($\al=1$) and $S$
($\al=2)$ being the inflaton and a ``stabilizer'' field
respectively. The Lagrangian density for $z^\al$'s within SUGRA
can be written as
\beqs \beq\label{Saction1}  {\cal  L} = \sqrt{-\mathfrak{g}}
\lf-\frac{1}{2}\rcc +K_{\al\bbet} \partial_\mu z^\al \partial^\mu
z^{*\bbet}-\Vf\rg, \eeq
where summation is taken over the scalar fields $z^\al$ and the
\Ka\ $\Khi$ obeys \Eref{kdef}. Also $\Vf$ is the F--term SUGRA
potential given by
\beq \Vf=e^{\Khi}\left(K^{\al\bbet}(D_\al W) (D^*_\bbet
W^*)-3{\vert W\vert^2}\right),\label{Vsugra} \eeq \eeqs
where $D_\al W=W_{,z^\al} +K_{,z^\al}W$ with $\Whi$ being the
superpotential. Therefore, the desired SUGRA embedding of our
models requires the determination of $K$ and $W$ and is processed
below along the lines of \cref{rube,unvr2}.

Our task can be facilitated, if we determine the inflationary
track by the constraints
\beq \label{inftr} \vevi{S}=\vevi{\Phi-\Phi^*}=0.\eeq
Note that we extend here the meaning of the symbol $\vevi{Q}$
(compared to that in \Sref{set}) to include also the leftmost
constraint in \Eref{inftr}.

If we postulate that $W$ is linear w.r.t $S$ (assuming, e.g., that
both have the same $R$ charge) the only surviving term in
\Eref{Vsugra} along the track in \Eref{inftr} is
\beq \label{1Vhio}\vevi{\Vf}=\vevi{e^{K}K^{SS^*}\,
|W_{,S}|^2}\,.\eeq
The last factor above can reproduce $\Vhi$ in \Eref{vhi} if
\beq \label{Wn} W=\ld
S\Phi^{n/2}\>\>\mbox{since}\>\>\vevi{|W_{,S}|^2}=\Vjhi.\eeq
%\begin{itemize} \item[{\sf\ftn  (i)}]\item[{\sf\ftn (ii)}]\end{itemize}
The considered below $K$'s do not allow for the imposition of
another symmetry which may assist to uniquely determine the form
of $W$ above. Therefore, taking in to account the imposed $R$
symmetry, $W$ may include at renormalizable level both terms in
\Eref{Wn} with $n=2$ and $n=4$ together with a term $SM^2$ --
where $M$ is a mass scale. For simplicity we assume that one of
the two $\phc$-dependent terms is dominant and $M\ll1$. We expect
that a mild hierarchy, of order $0.001$, in the relevant
coefficients of the terms $S\phc^{n/2}$ with $n=2$ and $n=4$ is
enough to isolate the impact of each one of them, as in a similar
set-up analyzed in \cref{epole}.

The crucial quantities for our scenario $J$ and $\Vhi$ in
\eqs{VJe}{vhi} can be achieved using $W$ in \Eref{Wn} for
judiciously chosen $K$'s. More specifically, the proposed $K$'s,
\tklmst, include two contributions without mixing between $\Phi$
and $S$, i.e.,
\beq
\tklmst=\tklm+\kst\>\>\mbox{with}\>\>l=1,2\>\>\mbox{and}\>\>\m=\e,\>\t
\label{ktot}\eeq
whereas the indices ``s'' and ``st'' are just descriptive, i.e.,
they do not take numerical or symbolic values as $l$ and \m. From
the contributions of $\tklmst$, $\kst$ successfully stabilizes $S$
along the path in \Eref{inftr} without invoking higher order
terms. We adopt the form \cite{su11}
\beq \kst=\nst\ln\lf1+{|S|^2/\nst}\rg\>\>\mbox{with}\>\>0<\nst<6,
\label{kst}\eeq
which parameterizes \cite{su11} the compact manifold $SU(2)/U(1)$
with curvature $2/\nb$. On the other hand, $\tklm$ should lead to
the same \Kaa metric with \klm\ in \Eref{kabm} and it should be
augmented by a shift symmetry which assures $\vevi{\tklmst}=0$. To
achieve these goals we set
\beqs\beq \tklm=\klm+K_{\rm
Msh}\>\>\mbox{with}\>\>l=1,2\>\>\mbox{and}\>\>\m=\e,\>\t\label{tks}\eeq
whereas the index ``sh'' is again descriptive. Here, $K_{\rm Msh}$
contains an holomorphic (and an anti-holomorphic) part which
fulfills the aforementioned prerequisites if it assumes the form
\beq K_{\rm Msh}=-(\na/2)\left(1-(\sqrt{2}\phc)^{\nm}\right)^{-p}-
(\na/2)\left(1-(\sqrt{2}\phc^{*})^{\nm}\right)^{-p},\label{kmsh}\eeq
which can be specified as follows
\beq K_{\rm Msh}=\bcs -(\na/2)\left(1-\sqrt{2}\phc\right)^{-p}-(\na/2)\left(1-\sqrt{2}\phc^{*}\right)^{-p}\>\>&\mbox{for \m=\e,}\\
-(\na/2)\left(1-2\phc^2\right)^{-p}-(\na/2)\left(1-2\phc^{*2}\right)^{-p}
&\mbox{for \m=\t.}\ecs\label{ksh}\eeq\eeqs
With these ingredients, we can easily confirm that
\beq
\vevi{e^K}=\vevi{K_{SS^*}}=1\>\>\mbox{and}\>\>\vevi{K_{\Phi\Phi^*}}=J^2\>\>\mbox{for}\>\>K=\tklmst.\label{ekss}\eeq
Upon substitution of the leftmost expression above together with
\Eref{Wn} into \Eref{1Vhio} we arrive at $\vevi{\Vf}=\Vhi$ and so
the inflationary setting based on \eqs{vhi}{VJe} readily emerges.

%. This result in conjunction with the rightmost output in
%\Eref{ekss} convinces that .

\renewcommand{\arraystretch}{1.3}
\begin{table}[t!]
\bec\begin{tabular}{|c|c|c|c|} \hline
{\sc Fields}&{\sc Eingestates} & \multicolumn{2}{|c|}{\sc Masses
Squared}\\\hline\hline
\multicolumn{4}{|c|}{Scalars}\\ \hline
%\what \what{m}^2_{\psi\pm}
$1$ real &$\what \th$ & $m^2_{\th}$&$6\Hhi^2$\\
$1$ complex&$S$ &$m^2_{S}$&
$(6/\nst+3n^2\fm^{p+2}/\nm^2pN\sg^{\nm}\fr)\Hhi^2$\\
\hline%
\multicolumn{4}{|c|}{Spinors}\\\hline
$2$ Weyl &$({{\psi}_{S}\pm
\what{\psi}_{\Phi})/\sqrt{2}}$&${m}^2_{\psi\pm}$&
$m^2_{S}-6\Hhi^2/\nst$ \\\hline
\end{tabular}\eec
\caption{\sl\small Mass-squared spectrum along the path in
\Eref{inftr} for $K=\tklmst$ with $l=1,2$ and $\m=\e,
\t$.}\label{tab2}
\end{table}
\renewcommand{\arraystretch}{1.}

To consolidate the SUGRA embedding of our models we check if the
configuration in \Eref{inftr} is stable w.r.t the excitations of
the non-inflaton fields $\what{\th}$ -- defined above \Eref{Hhi}
-- and $S$. In particular, we find the expressions of the masses
squared $m^2_{\chi^\al}$ (with $\chi^\al=\th$ and $S$) arranged in
\Tref{tab2}. These expressions assist us to appreciate the role of
$\nst$ with $0<\nst<6$ in retaining positive $m^2_{S}$ -- in
practise we use $\nst=1$. Also we confirm that $
m^2_{\chi^\al}\gg\Hhi^2=\Vhio/3$ for $\sgf\leq\sg\leq\sgx$. In
\Tref{tab1} we display the masses $m^2_{\psi^\pm}$ of the
corresponding fermions with $\psi_{S}$ and
$\what\psi_{\Phi}=\sqrt{K_{\Phi\Phi^*}}\psi_{\Phi}$ being the Weyl
spinors associated with $S$ and $\Phi$ respectively. Inserting the
derived mass spectrum in the well-known Coleman-Weinberg formula
we can find the one-loop radiative corrections, $\dV$ to $\Vhi$ --
cf. \cref{jhep}. It can be verified that our results are immune
from $\dV$, provided that the renormalization group mass scale
$Q$, is determined by requiring $\dV(\sgx)=0$ or $\dV(\sgf)=0$.
E.g., selecting the former condition for BPs \bpae, \bpbe, \bpat\
and \bpbt\ in \Tref{tab1} we obtain correspondingly
\beq Q=2.8\cdot10^{-10}, 1.6\cdot10^{-10},
7.6\cdot10^{-10}\>\>\mbox{and}\>\> 4.2\cdot10^{-10},
\label{qsugra} \eeq
using for definiteness $K=\kam$. Under these circumstances, our
results in the SUGRA set-up can be reproduced by using exclusively
the ingredients of \eqs{vhi}{VJe} as in the non-SUSY set-up.
Therefore, hereafter we do not discriminate our results for $\klm$
and $\tklmst$ with fixed $l$ and $\m$ since these are identical.

\section{Inflation Analysis} \label{ana}

The duration of the slow-roll \etpmi\ is determined by the
condition:
\beq{\ftn\sf
max}\{\eph(\phi),|\ith(\phi)|\}\leq1,\label{srcon}\eeq
where the slow-roll parameter $\eph$ takes a common form for any
$\klm$ -- with $l=1,2$ and \m=\e, \t\ as usual -- which is
\beqs \beq \label{eph} \epsilon=\left({\Ve_{\rm
I,\se}\over\sqrt{2}\Ve_{\rm I}}\right)^2=\frac{n^2
\fm^{p+2}}{\nm^2p N\sg^{\nm}\fr}\>\>\mbox{for}\>\>K=\klm,\eeq
whereas $\ith$ can be expressed by separate formulas for $\kam$
and $\kbm$ as follows
\beq\label{eta} \eta={\Ve_{\rm I,\se\se}\over\Ve_{\rm I}}\simeq
\frac{n\fm^{p+1}}{\nm pN\sg^{\nm}\fr}\cdot \bcs \lf2^{2-\nm}n
\fm\far-1-\sg^{\nm}-p\sg^{\nm}(p\sg^{\nm}+3)\rg/\far&\mbox{for $K=\kam$,}\\
2^{2-\nm}n \fm -2-p\sg^{\nm}&\mbox{for $K=\kbm$.}\ecs \eeq\eeqs
%p (n + p) s^4)
For the expressions above we employ $J$ in \Eref{VJe}, without
expressing explicitly $\Vhi$ in terms of $\se$ given in
\Eref{sesg}. Expanding $\fm$ for $\sg\ll1$ and neglecting terms of
order $\sg^{2\nm}$ or larger we obtain $\sg_{\eph\rm f}$ and
$\sg_{\ith\rm f}$ which saturate \Eref{srcon} for the $\eph$ and
$\ith$ criteria respectively. In practice, we mostly obtain
$\sg_{\eph\rm f}>\sg_{\ith\rm f}$ and so the only relevant values
are
\beq \sg_{\eph\rm f}\simeq\bcs \sqrt{n(n-1)/2(n^2+p(n-1)N}&\mbox{for $K=\kae$,}\\
2n\lf n(2+p)+\sqrt{4Np(1+4p)+n^2(p+2)^2}\rg^{-1}&\mbox{for $K=\kbe$,}\\
n/\sqrt{n^2 (2 + p)-4 N p}&\mbox{for $K=\kat$,}\\
\sqrt{2n}\lf n (2 + p) + \sqrt{16 N p (1 + p) + n^2 (2 +
p)^2}\rg^{-1/2}&\mbox{for $K=\kbt$.}\ecs \label{sgf}\eeq

The number of e-foldings $\Ns$ that the scale $\ks=0.05/{\rm Mpc}$
experiences during \etpmi\ can be computed using the standard
formula
\begin{align} \Ns=\int_{\sef}^{\sex} d\se\frac{\Vhi}{\Ve_{\rm
I,\se}}=2^{\nm-2}\frac{N}{n}\cdot\bcs{p\sgx^{\nm}}/{\fms^{p+1}} &\mbox{for $K=\kam$,}\\
\lf1+({\fbrs-1})/{\fms^{p+1}}\rg &\mbox{for $K=\kbm$,}\ecs
\label{Nhi}\end{align}
where $\sgx~[\sex]$ is the value of $\sg~[\se]$ when $\ks$ crosses
the inflationary horizon and we take in account that $\sgx\gg\sgf$
in the analytic expressions above. Hereafter, the variables with
subscript $\star$ are evaluated at $\sg=\sgx$ -- e.g.,
$\frs=\fr(\sgx)$. It is not doable to solve accurately the
analytic expressions in \Eref{Nhi} w.r.t $\sgx$ in order to pursue
our analytic approach. Nonetheless, if we set $\sgx^{\nm}\simeq1$
in their numerators, overestimating somehow $\Ns$, then the
solution w.r.t $\sgx$ becomes feasible since $\Ns$ gets simplified
as follows
\begin{equation}
\label{sgx} \Ns\simeq2^{\nm-2}\frac{N}{n}\cdot\bcs{p}/{\fms^{p+1}} &\mbox{for $K=\kam$,}\\
(1+{p}/{\fms^{p+1}}) &\mbox{for $K=\kbm$}\ecs
\>\>\Rightarrow\>\>\sgx\simeq \lf1 - \lf \frac{\nm pN}{2n
\Ns}\rg^{\frac{1}{p+1}}\rg^{2^{1-\nm}}
\end{equation}
for any $\klm$ in \Eref{kabm}. Note that for $K=\kbm$ we further
assumed that $\Ns\gg N/\nm n$, which is confirmed a posteriori. It
is clear that we need to set $\sgx<1$ by construction to obtain a
large enough $\Ns$ -- see \Tref{tab1}. This fact assists us to
avoid any destabilization of our inflationary scenario due to
higher order non-renormalizable terms in $W$ and/or \klm -- see
\eqs{Wn}{tks}. Taking also into account that
$\Vhi(\sgx)^{1/4}\leq1$ -- see \Fref{fig2} --, we expect
\cite{lindeg} that corrections from quantum gravity may be in
principle under control. However, it is premature to assess if
this naive expectation remains valid within a consistent
string-theoretical realization. This task is implemented for the
$\alpha$ attractors in \cref{fibre} where the importance of the
quantum-gravity effects are explicitly evaluated in the context of
fibre inflation \cite{fibre1}.

The amplitude $\As$ of the power spectrum of the curvature
perturbations generated by $\sg$ can be calculated at $\sg=\sgx$
as a function of $\ld$. With given $\As$ we can also derive $\ld$
as follows
\beq \label{lan} \As= \frac{1}{12\pi^2} \; \frac{\Ve_{\rm
I}^{3}(\sex)}{|\Ve_{\rm I,\se}(\sex)|^2}\>\>\Rightarrow\>\>
\ld\simeq2^{\frac{5-2\nm}{2}+\frac{n}{4}}\frac{\sqrt{3\As}n\pi\fms^{p/2+1}}{
\sqrt{pN\sgx^{n+\nm}\frs}}\>\>\mbox{for}\>\>K=\klm, \eeq
where $\sgx$ is given by \Eref{sgx}. We did not replaced it in the
expression above to avoid the exposition of the lengthly final
result.

The remaining inflationary observables (i.e. \ns, its running
$\as$ and the tensor-to-scalar ratio $r$) are calculated via the
relations
\beqs\baq \label{ns} && \ns=\: 1-6\epsilon_\star\ +\
2\eta_\star\simeq 1-\frac{p+2}{(p+1)\Ns},\\
&& \label{rs} r=16\epsilon_\star\simeq
2^{{(p(4-\nm)+2)}/(p+1)}\frac{(pn^pN)^{1/(p+1)}}{(p+1)N_\star^{(p+2)/(p+1)}},\\&&
\label{as} \as =\:{2\over3}\left(4\eta_\star^2-(n_{\rm
s}-1)^2\right)-2\xi_\star\simeq-\frac{p+2}{(p+1)N^2_\star},\eaq\eeqs
where $\xi={\Ve_{\rm I,\widehat\sg} \Ve_{\rm
I,\widehat\sg\widehat\sg\widehat\sg}/\Ve_{\rm I}^2}$ and the
approximate expressions are obtained by expanding the exact result
for all possible $\klm$ in powers of $1/\Ns$ and keeping the
lowest order term. The common expressions for $\eph\gg\eta$ and
$\sgx$ in \eqs{eph}{sgx} justify the ``unified'' results for all
possible $\klm$. If we set ${\sf p}=p+2$ we can see that the \ns\
expression in \Eref{fbp} is readily reproduced despite the fact
that the kinetic mixing therein deviates from that in \Eref{VJe}.
Note that $\ns$ and $\as$ are independent of $N$ at the lowest
order. This feature, however, is not totally preserved as we show
in our numerical results below -- in accordance with the behavior
of similar models \cite{terada}. On the other hand, the
independence of $\ns, r$ and $\as$ from $n$ (and $\nm$ in a
considerable degree) clearly defends the existence of an attractor
behavior which is influenced only by the parameters $N$ and $p$.

\section{Numerical Results}\label{num}

Our analytic findings above can be checked and refined
numerically. At first, we confront the quantities in \Eref{Nhi}
with the observational requirements \cite{act}
\beq \Ns \simeq61.3+\frac{1-3w_{\rm rh}}{12(1+w_{\rm
rh})}\ln\frac{\pi^2g_{\rm rh*}\Trh^4}{30\Vhi(\sgf)}+
\frac14\ln{\Vhi(\sgx)^2\over g_{\rm
rh*}^{1/3}\Vhi(\sgf)}\>\>\>\mbox{and}\>\>\>\As\simeq2.132\cdot10^{-9},\label{Prob}\eeq
%}{
where we assume that \etpmi\ is followed in turn by an oscillatory
phase with mean equation-of-state parameter $w_{\rm rh}$ --
$w_{\rm rh}\simeq0$ for $n=2$ and $w_{\rm rh}\simeq1/3$ for $n=4$
-- radiation and matter domination. Motivated by implementations
\cite{tmhi} of non-thermal leptogenesis, which may follow \etpmi,
we set $\Trh\simeq10^9~\GeV$ for the reheat temperature. Although
not crucial for the resulting magnitude of $\Ns$ we mention that
we take for the energy-density effective number of degrees of
freedom $g_{\rm rh*}=228.75$ inspired by the MSSM spectrum.

Enforcing \Eref{Prob} we can restrict $\ld$ and $\sgx$ -- see
\Eref{Nhi}. In general, we obtain $\ld\simeq(0.5-20)\cdot10^{-5}$
in agreement with \Eref{lan}. Regarding $\sgx$ we assume that
$\sg$ starts its slow roll below the location of kinetic pole,
i.e., $\sg=1$, consistently with our approach to SUGRA as an
effective theory below $\mP=1$. The closer to unity $\sgx$ is set,
the larger $\Ns$ is obtained. Therefore, a tuning of the initial
conditions is required which can be somehow quantified defining
the quantity
\beq \Dex=\left(1 - \sgx\right).\label{dex}\eeq
The naturalness of the attainment of \etpmi\ increases with
$\Dex$.

\begin{figure}[!t]\centering\hspace*{-0.3cm}
\includegraphics[width=60mm,angle=-90]{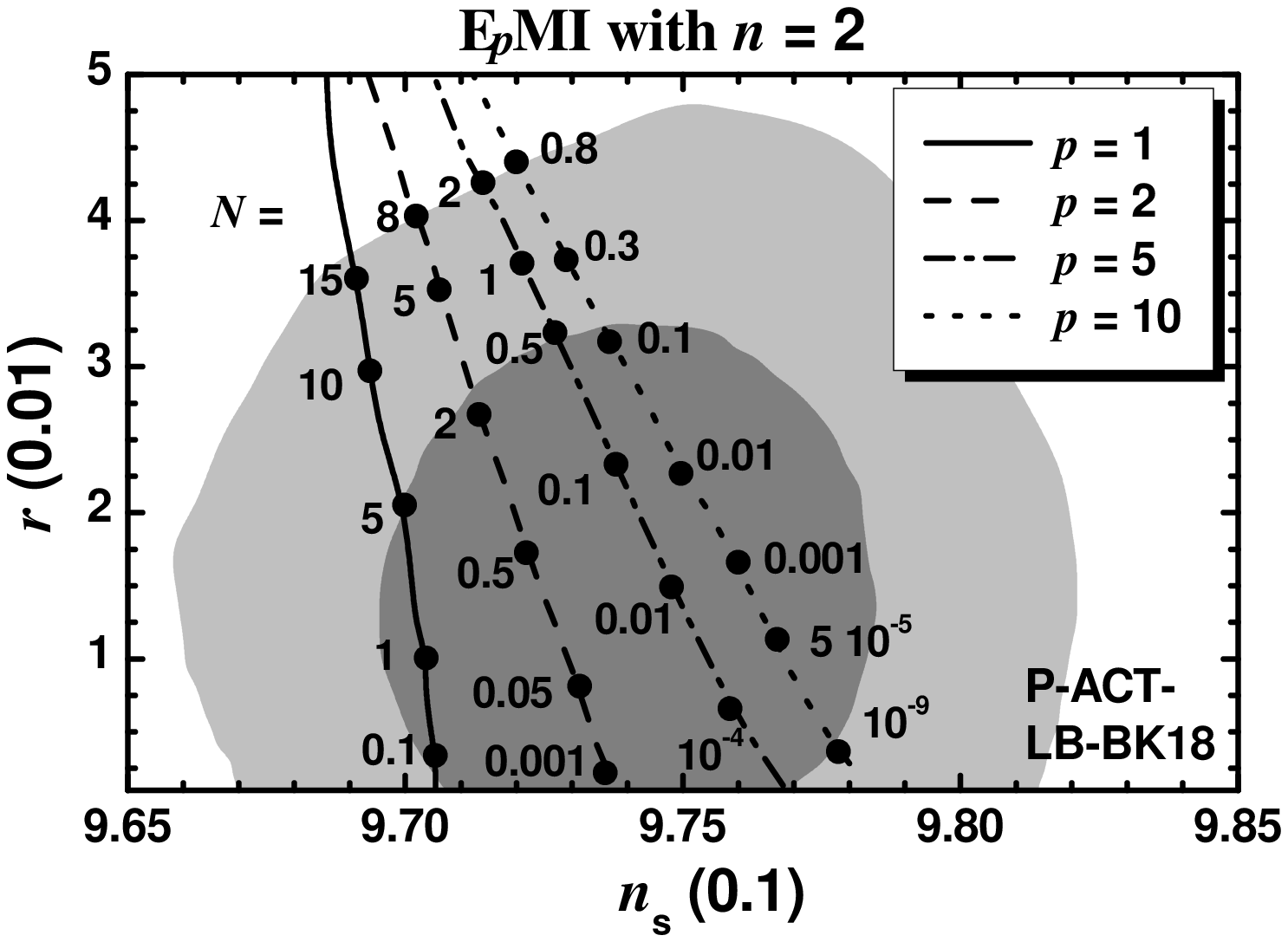}\hspace*{-0.7cm}
\includegraphics[width=60mm,angle=-90]{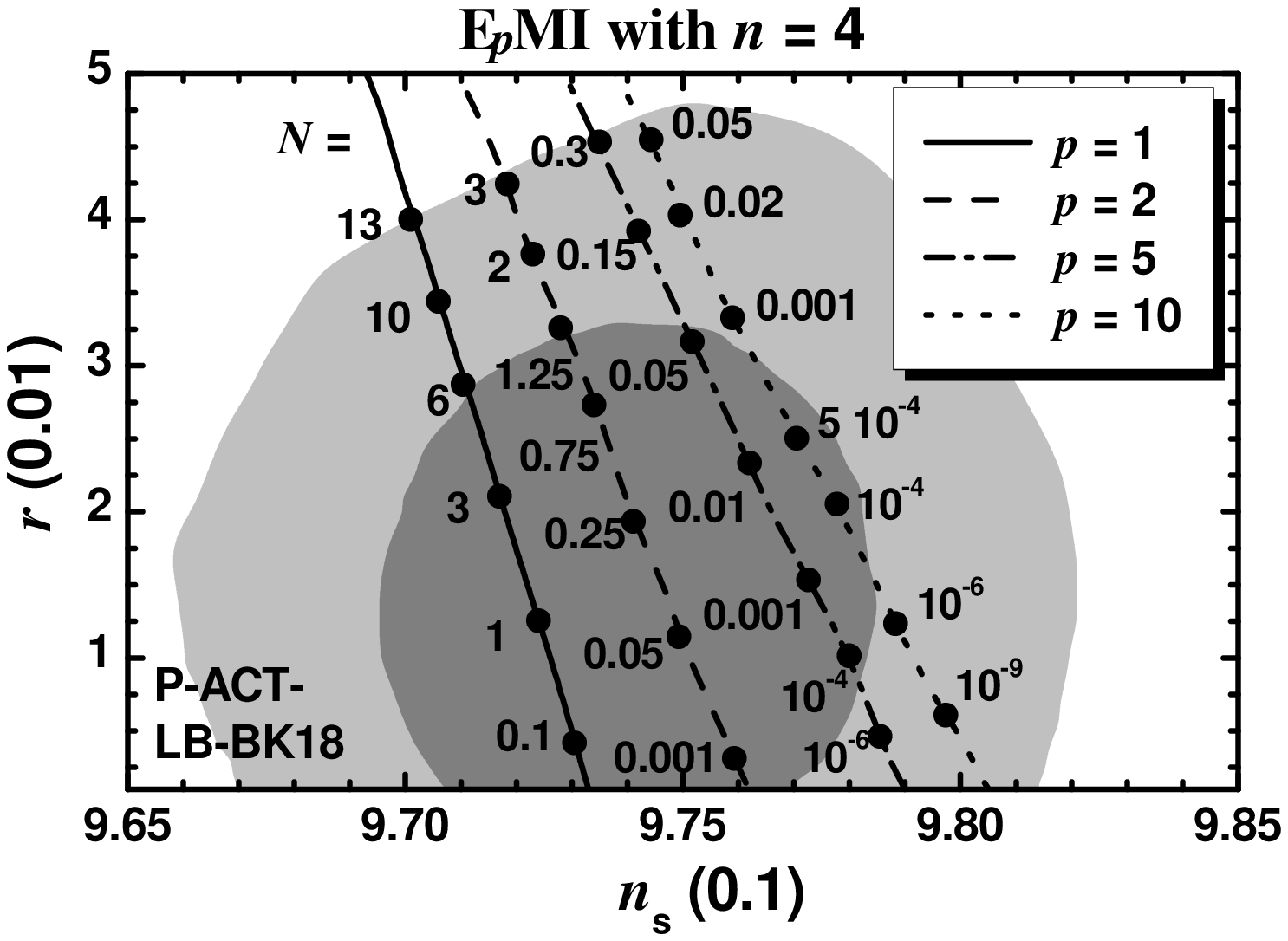} \\\hspace*{-0.3cm}
\includegraphics[width=60mm,angle=-90]{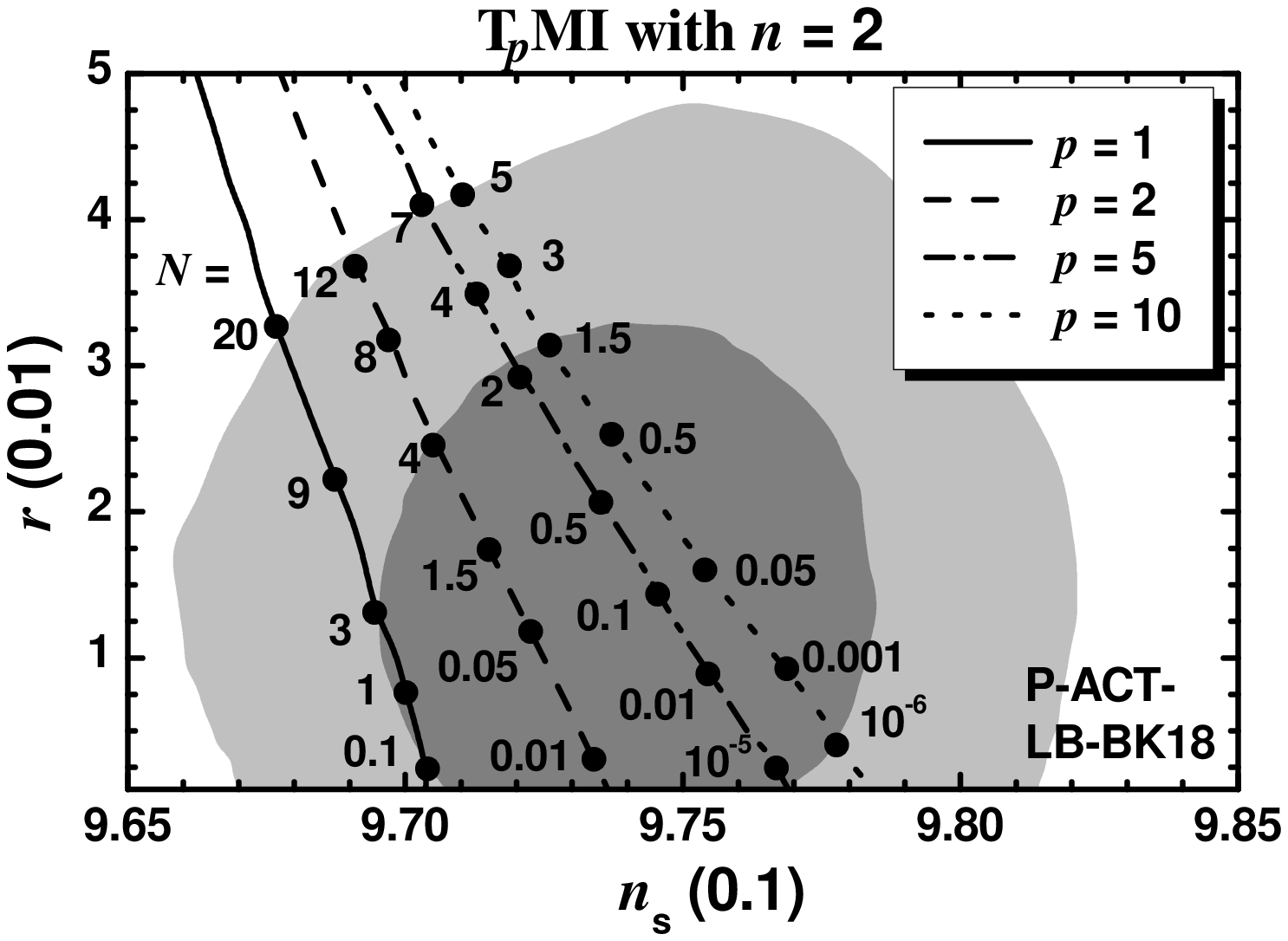}\hspace*{-0.7cm}
\includegraphics[width=60mm,angle=-90]{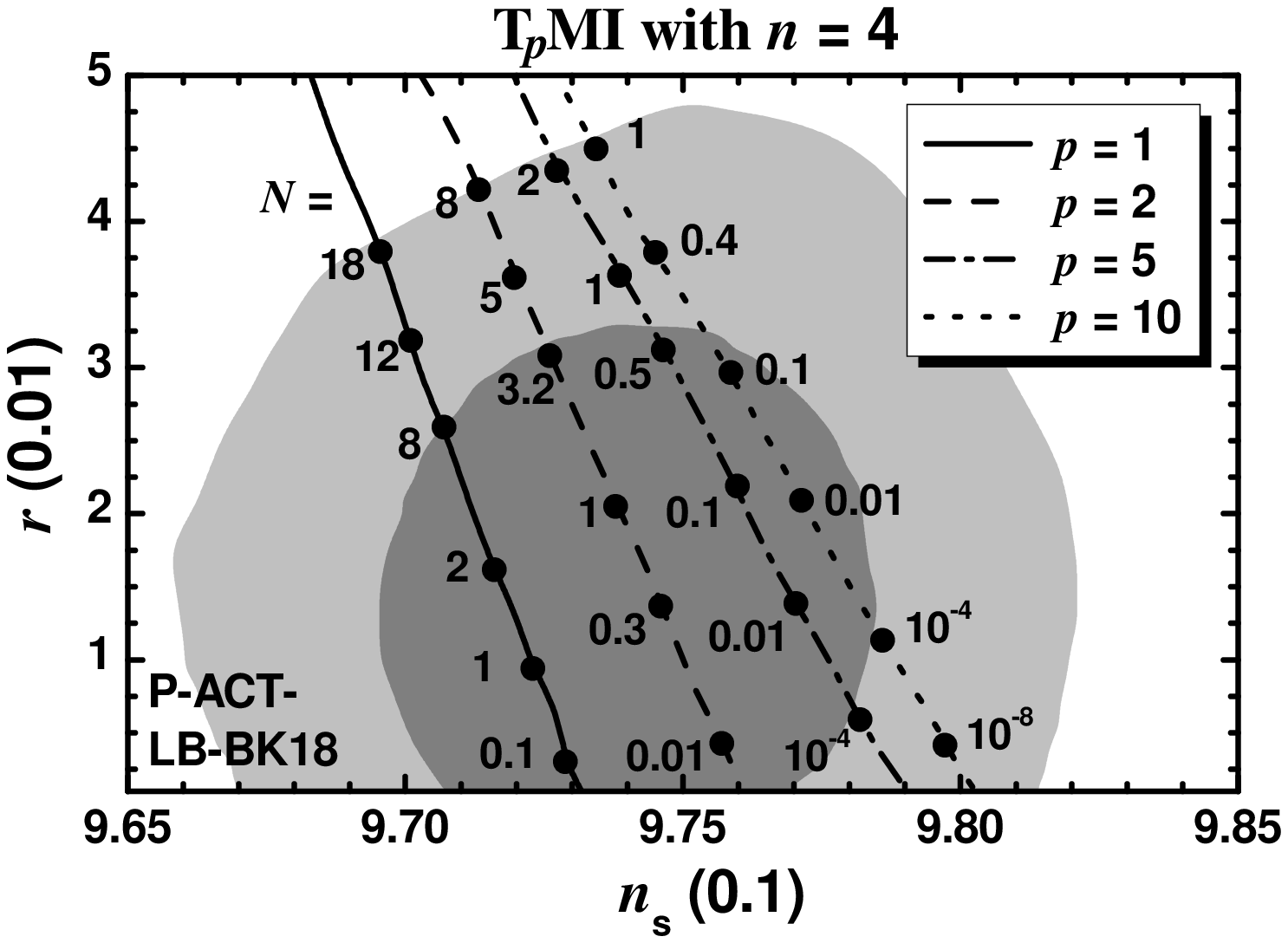}
\caption{\sl\small Allowed curves in the $\ns-r$ plane for \epmi\
(upper plots) and \tpmi\ (lower plots) with $n=2$ (left panels) or
$4$ (right panels), various $p$ values -- shown in the legends --
and  $N$ values indicated on the curves. The marginalized joint
$68\%$ [$95\%$] c.l. regions from \actc\ data are depicted by the
dark [light] shaded contours.}\label{fig3}
\end{figure}

After the extraction of $\ld$ and $\sgx$, we compute the models'
predictions via the definitive expressions in \eqss{ns}{rs}{as},
for any selected set $(n,N,p)$. Our outputs are encoded as lines
in the $\ns-r$ plane superposed on the $68\%$ and $95\%$ c.l.
regions from \actc\ data \cite{actin} depicted by the dark and
light shaded contours respectively -- see Fig.~\ref{fig3}. The two
upper plots are devoted to \epmi\ whereas the two lower plots are
for \tpmi. In the left plots we fix $n=2$ whereas in the two right
plots $n=4$. We draw solid, dashed, dot-dashed and dotted lines
for $p=1, 2, 5$ and $10$ respectively and show the variation of
$N$ along each line. It is evident that the whole observationally
favored range in the $\ns-r$ plane is covered varying $p$ and $N$.
In particular, as inferred from \Eref{ns}, $\ns$ increases with
$p$ regarding its value in \Eref{nsnmi}, and renders \etpmi\
excellently compatible with data. Moreover, this increase becomes
slower for large $p$'s and so the upper bound on $\ns$ in
\Eref{data} is not saturated for realistically large $p$ values.
For this reason we set by hand an upper bound $p=10$.

From Fig.~\ref{fig3}, it is also deduced that as $n$ increases
above $2$ the allowed curves move to the right. This behavior can
be understood from the fact that the $\Ns$ required by the
leftmost condition in \Eref{Prob} is higher (about $57$) for $n=4$
than for $n=2$ (about $51$) -- see below. If we compare the two
upper with the two lower plots of \Fref{fig3} with constant $n$ we
can remark that the inclination of the various lines increases for
\tpmi\ relatively to that in \epmi. This fact declares some minor
$\nm$ dependence appeared in the computation of $r$ -- see
\Eref{rs}. This effect is disadvantageous for $n=2$ but beneficial
for $n=4$. Therefore, we could say that \epmi\ fits better with
$n=2$ whereas \tpmi\ with $n=4$. These minor numerical variations,
encountered also within \etmi, do not invalidate the attractor
behavior arisen in our models regarding the essential independence
of the observables ($\ns, r$ and $\as$) from the parameters $n$
and $\nm$.

Another remark from Fig.~\ref{fig3} is that the $N$ values for
\epmi\ turn out to be suppressed compared to the values used in
\tpmi, especially for large $p$'s. This result may be attributed
to the factor $2^{p/(p+1)}\simeq(1.1-1.9)$ for $0.1\leq p\leq10$
which slightly increases $r$ within \epmi\ compared to its value
in \tpmi -- see \Eref{rs}. As a consequence, lower $N$ values are
necessitated for the attainment of observationally acceptable $r$
values according to \Eref{data}. E.g., for $(n,p)=(4,2)$ we obtain
\beq 8\cdot10^{-4}\lesssim N\lesssim3 \>\>\mbox{for \epmi\
and}\>\>7\cdot10^{-3}\lesssim N\lesssim8 \>\>\mbox{for
\tpmi.}\>\>\nonumber\eeq
This fact signals some tuning which can be avoided selecting low
$p$ values especially in \epmi. It is worth noticing, lastly, that
natural $N$ values result to $r$ values which lie within the reach
of planned experiments \cite{det} aiming to discover primordial
gravitational waves.

%\epmi\ and \tpmi\ we can see that

\begin{figure}[!t] \centering\hspace*{-0.3cm}
\includegraphics[width=60mm,angle=-90]{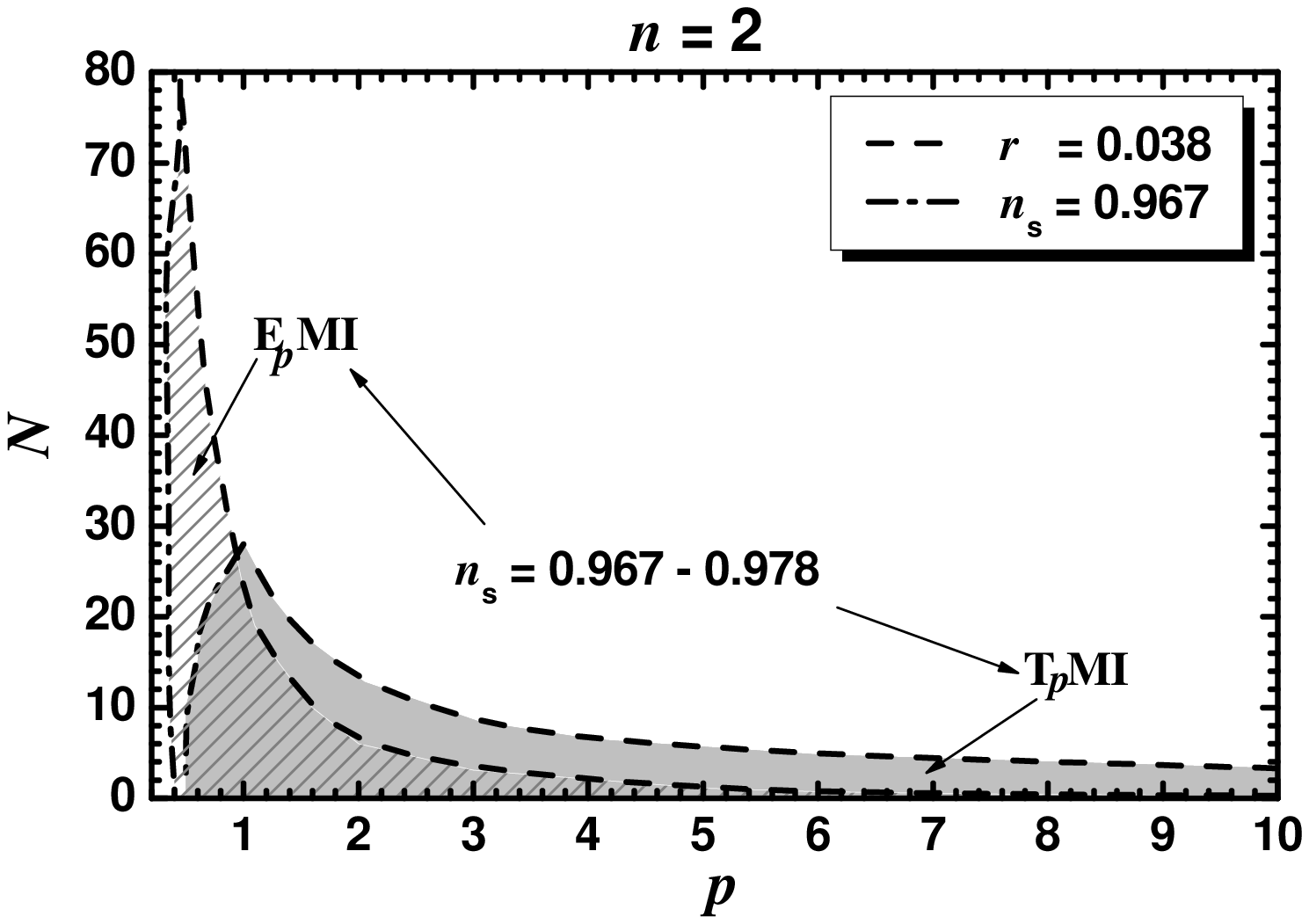}\hspace*{-0.7cm}
\includegraphics[width=60mm,angle=-90]{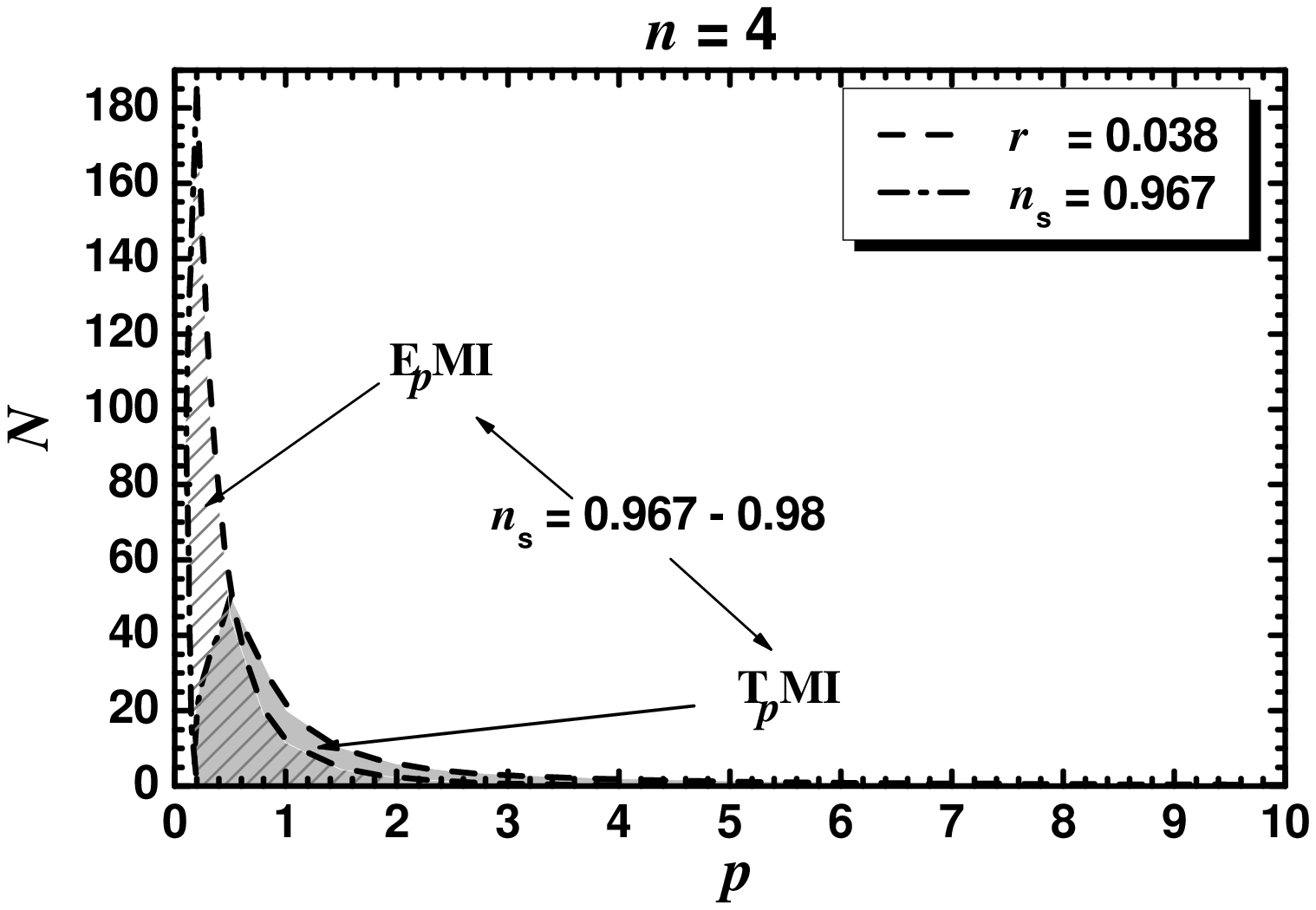}
\caption{\sl\small  Allowed (shaded for \tpmi\ and hatched for
\epmi) regions as determined by \eqs{data}{Prob} in the $p-N$
plane for $n=2$ (left panel) or $n=4$ (right panel). The
conventions adopted for the boundary lines are also
shown.}\label{fig4}
\end{figure}

Our plots in \Fref{fig3} reveal that $\ns$ slightly decreases as
$N$ increases and so an lower bound on $N$ can be obtained from
the lower bound on $\ns$ in \Eref{data}. On the other hand, $N$ is
also bounded from above from the bound on $r$ in \Eref{data} since
$r$ increases with $N$. Namely, the maximal allowed $N$ from the
last constraint decreases as $p$ increases. The competition of
both restrictions on $N$ is shown in \Fref{fig4} where we
delineate the allowed regions (shaded for \tpmi\ and lined for
\epmi) in $p-N$ plane for $n=2$ (left panel) and $n=4$ (right
panel). We see that in both cases \epmi\ allows for larger and
requires lower maximal $N$ values $N_{\rm max}$ depending on $p$.
Summarizing our results for $n=2$ -- see left plot of \Fref{fig4}
-- we arrive at the following allowed ranges
\beqs\begin{align}\label{res2} 0.3\lesssim N_{\rm
max}\lesssim79\>\>\mbox{and}\>\>0.2\lesssim\Dex/100\lesssim61.4\>\>
\mbox{or}\>\>3.3\lesssim N_{\rm
max}\lesssim28\>\>\mbox{and}\>\>1.6\lesssim\Dex/100\lesssim53.8\end{align}
%dictates
for \epmi\ or \tpmi, respectively. Also $\as\simeq
-(4.2-6.5)\cdot10^{-4}$ and $\Ns\simeq(50.5-52.1)$ for both
\etpmi. On the other hand, for $n=4$ -- see right plot of
\Fref{fig4} -- we obtain
\begin{align} \label{res4} 0.015\lesssim N_{\rm
max}\lesssim185\>\>\mbox{and}\>\>0.7\lesssim\Dex/100\lesssim45.3\>\>
\mbox{or}\>\>0.4\lesssim N_{\rm
max}\lesssim50\>\>\mbox{and}\>\>0.3\lesssim\Dex/100\lesssim39\end{align}\eeqs
for \epmi\ or \tpmi, respectively. Also $\as\simeq
-(3.6-5.6)\cdot10^{-4}$ and $\Ns\simeq(55.6-57.9)$ for both
\etpmi. In both \eqs{res2}{res4} the maximal $\ns$ values are
obtained for the lowest possible $N$ and the largest $p$ values
whereas the minimal $\ns$  values are achieved at the lowest $p$
values independently of $N$. As regards $\Dex$, its maximal values
are obtained for the largest $p$'s and $N$'s whereas its minimal
values for the lowest $p$'s and $N$'s. It is notable that the
maximal $\Dex$ values are much larger than the ones derived within
\etmi\ as those are exposed in \cref{epole, sor} and therefore,
the present models can be characterized as more natural from the
point of view of the tuning in the initial conditions.

%%%%%%%%%%%%%%%%%%%%%%%%%%%%%%%%%%%%%%%%%%%%%%%%%%%%%%%%%%%%%%%%%%%%
%%%%%%%%%%%%%%%%%%%%%%%%%%%%%%%%%%%%%%%%(or $\tkamst$ and $\tkbmst$)

We complete our analysis by checking the accuracy of our
analytical expressions and the equivalence between $\kam$ and
$\kbm$ for fixed \m. For both aims we use as reference the four
BPs of \Tref{tab1} where we display results -- by using the $K$'s
in \Eref{kabe} for \epmi\ and in \Eref{kab} for \tpmi\ -- derived
by our numerical code and by our approximate expressions in
\Sref{ana}. Note that the analytical results are indistinguishable
for \kam\ and \kbm\ with fixed \m\ as regards \ns, \as, $r$,
$\sgx$ and $\Dex$ since those quantities are the same for both
sets of $K$'s -- see \Sref{ana}. On the contrary, $\ld$ and $\sgf$
are estimated with different formulas -- see \eqs{sgf}{lan} -- for
each \klm. Despite their simplicity, our analytic results
approximate rather well the precise numerical values. The
approximation becomes more accurate as $\Ns$ in \Eref{sgx}
approximates better its numerical value in \Eref{Nhi}. E.g., for
\bpbt\ where $\sgx\ll1$, the analytic $\Ns$ and $\ld$ values
deviate substantially from their precise numerical values.
However, $\sgx, \ns, \as$ and $r$ remain accurate enough. Also,
from \Tref{tab1} it is evident that interchanging $K=\kam$ with
$K=\kbm$ with fixed \m\ for the same BP does not influence
essentially the results. Just for definiteness we mention that
\Fref{fig3} and \ref{fig4} were constructed by considering
$K=\kbe$ and $\kat$.

%as $N$ is kept low

%\newpage

\section{Conclusions}\label{con}

Prompted by the \actc\ data which seems to disfavor the
conventional \etmi\ (i.e., E- and T-model inflation) at more than
two standard deviations, we proposed two variants called \etpmi\
which assure compatibility with the current observations at the
cost of just one more parameter $p$ which strengthens the
operation of the pole, already present in E/TMI, by an extra
exponent $(p+2)$ -- see \Eref{VJe}. As a result $\ns$ approaches
its values in \Eref{ns} in nice agreement with \Eref{data}. This
convergence of both proposed models towards the same more or less
values of observables allows us to identify the emergence of a new
type of attractors named $(N,p)$ attractors.

More specifically, our setting was established as a non-linear
sigma model using the potential and the \Ka s in \eqs{vsg}{kabm}.
It can be implemented in SUGRA too, employing the super- and \Ka s
given in \eqs{Wn}{ktot}. Both formulations yield the potential and
the kinetic mixing in \eqs{vhi}{VJe}. Confining $(N,p)$ in the
ranges of \Fref{fig4} we achieved a nice covering of the present
data -- its compatibility especially with the $n=4$ case for low
$p$ and $N$ values is really impressive, as shown in \Fref{fig3}.
The resulting $r$ values could be accessible in the near future.
Since our models exhibit a mechanism for enhancing the canonical
normalized inflaton w.r.t the original one -- see \Fref{fig1} --
our solutions can be attained with subplanckian values of the
initial inflaton. It is gratifying that these solutions can be
studied analytically and rather accurately in a sizable portion of
the allowed parameter space. Namely, our final analytical outputs
are summarized in \eqss{ns}{rs}{as}.

As a bottom line, we introduced two new versions of pole
inflation, \etpmi, with robust predictions toward the ``sweet''
spot of the recent ACT data which signal the existence of an
attractor mechanism. Within our scheme, the reheating phase is not
constrained as, e.g., in \cref{rha,rhb,rhc,rhd, act5, r2drees} and
corrections to the inflationary potential from other sectors of
the theory are not required as, e.g., in \cref{act6,oxf,warm}.
Finally, our models -- especially \tpmi\ -- can be extended for a
gauge non-singlet Higgs superfield as done in \cref{jhep,sor} and
can be applied for non-SUSY models of hybrid inflation as done in
\cref{ethi}.

%the corresponding effective theory respects  up to $\mP$.

%\vspace*{-.5cm}

%\paragraph*{\small\bfseries\scshape Acknowledgments} {\small }

%\newpage

%\newcommand\jcap[3]{{\sl J.\ Cosmol.\ Astropart.\ Phys.\ }{\bf #1}, #3 (#2)}
%\newcommand\jcapn[4]{{\sl J.\ Cosmol.\ Astropart.\ Phys.\ }{\bf #1}, #3, no.~#4 (#2)}
%\newcommand\njp[3]{{\sl New.\ J.\ Phys.\ }{\bf #1}, #3 (#2)}

\def\prdn#1#2#3#4{{\sl Phys. Rev. D }{\bf #1}, no. #4, #3 (#2)}
\def\jcapn#1#2#3#4{{\sl J. Cosmol. Astropart.
Phys. }{\bf #1}, no. #4, #3 (#2)}

\end{document}